\documentclass[twocolumn]{aastex631}
\usepackage{amsmath,amssymb}
\usepackage{newtxmath}
\usepackage{comment}
\usepackage{soul} 
\received{June 19, 2023}
\revised{September 8, 2023}

\begin{document}

\title{Impact of Galaxy Clusters on Propagation of Ultra-High-Energy Cosmic Rays}

\author[0000-0001-5681-0086]{Antonio Condorelli}
\affiliation{Université Paris-Saclay, CNRS/IN2P3, IJCLab, 91405 Orsay, France}

\author[0000-0002-4202-8939]{Jonathan Biteau}
\affiliation{Université Paris-Saclay, CNRS/IN2P3, IJCLab, 91405 Orsay, France}
\affiliation{Institut universitaire de France (IUF)}

\author{Remi Adam}
\affiliation{Université Côte d'Azur, Observatoire de la Côte d'Azur, CNRS, Laboratoire Lagrange, France}



\begin{abstract}
Galaxy clusters are the universe's largest objects in the universe kept together by gravity. Most of their baryonic content is made of a magnetized diffuse plasma. We investigate the impact of such magnetized environment on ultra-high-energy-cosmic-ray (UHECR) propagation. The intracluster medium is described according to the self-similar assumption, in which the gas density and pressure profiles are fully determined by the cluster mass and redshift. The magnetic field is scaled to the thermal components of the intracluster medium under different assumptions. We model the propagation of UHECRs in the intracluster medium using a modified version of the Monte Carlo code {\it SimProp}, where hadronic processes and diffusion in the turbulent magnetic field are implemented. We provide a universal parametrization that approximates the UHECR fluxes escaping from the environment as a function of the most relevant quantities, such as the mass of the cluster, the position of the source with respect to the center of the cluster and the nature of the accelerated particles. We show that galaxy clusters are an opaque environment especially for UHECR nuclei. The role of the most massive nearby clusters in the context of the emerging UHECR astronomy is finally discussed.

\end{abstract}



\section{Introduction} \label{sec:intro}
Even though their existence has been known for more than a century, the nature and origin of cosmic rays at the highest energies remains elusive. Observations have allowed us to explore their spectral behavior and composition in terms of atomic mass on Earth \citep{Coleman:2022abf}, but the sources of ultra-high-energy cosmic rays (UHECRs){, i.e.\ cosmic rays above $10^{18}\ \rm eV$}, still remain unknown.

Rapid progress in computational high-energy astrophysics has dramatically advanced the study of acceleration mechanisms in systems ranging from the jets of stellar-sized objects such as gamma-ray bursts \citep[GRBs, ][]{Sudilovsky_2013} to the large-scale shocks surrounding galaxy clusters \citep{1995ApJ...454...60N,Ryu_2003,Kang_1997}. Galaxy clusters are the largest virialized structures, having typical radii $R_{\rm cl} = 1-2$\,Mpc and total masses $M \simeq 10^{14}-10^{15} M_{\odot}$, including both baryonic and dark matter. 
Strong turbulent magnetic fields, with root mean square values $B \simeq$ few $\mu$G, are present inside 
clusters, having typical coherence lengths of $5–30$\,kpc \citep{Bonafede_2010, Donnert_2018}. 
This implies that cosmic rays accelerated in candidate sources inside the clusters, e.g.\ by hypernovae or GRBs in star-forming galaxies, or in the accretion shocks, jets and radiolobes of active galactic nuclei (AGNs), can be confined for long times within clusters. They can undergo interactions with the enhanced baryonic content of the intracluster medium, whose profile is determined by Bremsstrahlung emission in X rays \citep{RevModPhys.58.1}.

Upper limits on the flux of neutrinos and gamma rays at ultra-high energies rule out a dominant origin  of UHECRs from exotic particles \citep{PhysRevLett.130.061001,Abreu_2022_photons}, which should then originate from extragalactic astrophysical sources. An extragalactic origin is corroborated by the observation of a dipolar anisotropy above 8\,EeV \citep{Aab_2018} and an evidenced correlation of UHECRs above 40\,EeV with extragalactic objects in the nearby universe \citep{Aab_2018_SS,Abreu_2022}. Some of these extragalactic sources could be hosted or shadowed by clusters. 

UHECR propagation in a specific cluster (e.g.\ the Virgo cluster) has been already treated in different works \citep{Dolag:2008py, Kotera_2009, Harari_2016, Fang_2018}. 
Although some of these theoretical works suggested that galaxy clusters are efficient UHECR calorimeters,  some authors recently claimed excesses of UHECRs from these structures \citep{Ding_2021, TelescopeArray:2021dfb}. Revisiting the  propagation of UHECRs in galaxy clusters is thus a timely topic. In the following, we evaluate whether UHECRs can escape from such environments and how clusters should be accounted for in UHECR astronomy. We provide in particular a single parametrization of the escaping flux, which depends on the mass of the cluster and on the UHECR features, such as energy and atomic mass.

The paper is organized as follows: we introduce the relevant properties of galaxy clusters and detail the way we compute the most important macroscopic quantities for our study in Section~\ref{Sec: 2-Galaxy cluster}; the microphysics of UHECR propagation in such environments is detailed in Section~\ref{Sec 4: Propagation}; we present our results and discuss the impact of our assumptions  in Section~\ref{Sec: 5-Param}. We finally draw our conclusions in Section~\ref{Sec: 9 Conclusion}.

\section{Intracluster medium modeling} \label{Sec: 2-Galaxy cluster}

Clusters of galaxies and the filaments that connect them are the largest structures in the present universe in which the gravitational force due to the matter over-density overcomes the expansion of the universe. Massive clusters have typical total masses of the order of $10^{15} \ M_{\odot}$, mostly in the form of dark matter ($70-80$\,\% of the total mass), while baryonic matter is harbored by galaxies (few \%) and composes the hot ($T \sim 10^{8}$\,K) and tenuous ($n_{\rm gas} \simeq 10^{-1} -10^{-4} \, \rm cm^{-3}$) gas ($15-20$\,\%) that forms the intracluster medium
\citep[ICM, ][]{VOIT_2005}. To model UHECR propagation in this environment, we need estimates of the gas density profile, of the magnetic field profile and of the coherence length. While the gas density is well understood and routinely derived from X-ray observations, this is not the case for the magnetic field, for which only a handful of measurements -- sometimes model-dependent -- are available in the literature \citep{Vacca}. From theoretical arguments, however, the magnetic-field strength is often assumed to scale with the ICM thermal density or pressure.

\subsection{Density profile}

An interesting feature of galaxy clusters is that they are self-similar objects at first order, so that their physical properties can be fully described given their mass and redshift \citep{KAISER_1986}. For instance, their universal pressure profiles (UPP) and universal density profiles (UDP) are now well constrained from observations \citep[e.g.,][]{2010A&A...517A..92A,Pratt:2022nrs}. Following \cite{2010A&A...517A..92A}, we use the UPP expressed as 
\begin{equation}
P(x) = \frac{P_0  \times P_{500}(M_{500},z) \times f(M_{500},z)}{(c_{500} x)^{\gamma_{\rm UPP}} \cdot (1+(c_{500} x)^{\alpha_{\rm UPP}})^{\frac{\beta_{\rm UPP} - \gamma_{\rm UPP}}{\alpha_{\rm UPP}}}},
\label{eq:UPP}
\end{equation}
with $P_{500}(M_{500},z)$ the self-similar normalization \citep{NAGAI_2007}, $f(M_{500},z)$ a small mass-dependence correction, and where $P_0$, $c_{500}$, $\alpha_{\rm UPP}$, $\beta_{\rm UPP}$, $\gamma_{\rm UPP}$ are parameters that describe the shape of the profile as a function of the scaled radius $x=r/R_{500}$.\footnote{The mass $M_{500}$ is defined within $R_{500}$, the radius within which the cluster density is 500 times the critical density of the universe at the cluster redshift.}

Similarly, we use the UDP as measured by \cite{Pratt:2022nrs}, which can be expressed as 
\begin{equation}
n(x) = \frac{A(M_{500}, z) \times f_0}{\left(x/x_{s}\right) \left(1 + (x/x_{s})^{\gamma_{\rm UDP}}\right)^{\frac{3 \beta_{\rm UDP} - \alpha_{\rm UDP}}{\gamma_{\rm UDP}}}}.
\label{eq: UDP}
\end{equation}
The quantity $A(M_{500}, z)$ describes the normalization as a function of mass and redshift, and the parameters $f_0$, $x_s$, $\alpha_{\rm UDP}$, $\beta_{\rm UDP}$, $\gamma_{\rm UDP}$ describe the shape \citep[see also][for another calibration of the UDP]{Ghirardini_2019}. 

The gas density and pressure profile are expected to be connected. This provides us with an alternative way to describe the ICM thermal density given the pressure profile. Assuming a polytropic relation between gas density and pressure, using a sample of massive nearby clusters, \cite{Ghirardini_2019b} measured
\begin{equation}
     P(x) = C \times n(x)^{k},
     \label{eq:ploytropic_DP}
\end{equation}
where $k = 1.19$ and $C$ is a normalization constant. 

\begin{figure}
\centering
\includegraphics[width=\columnwidth]{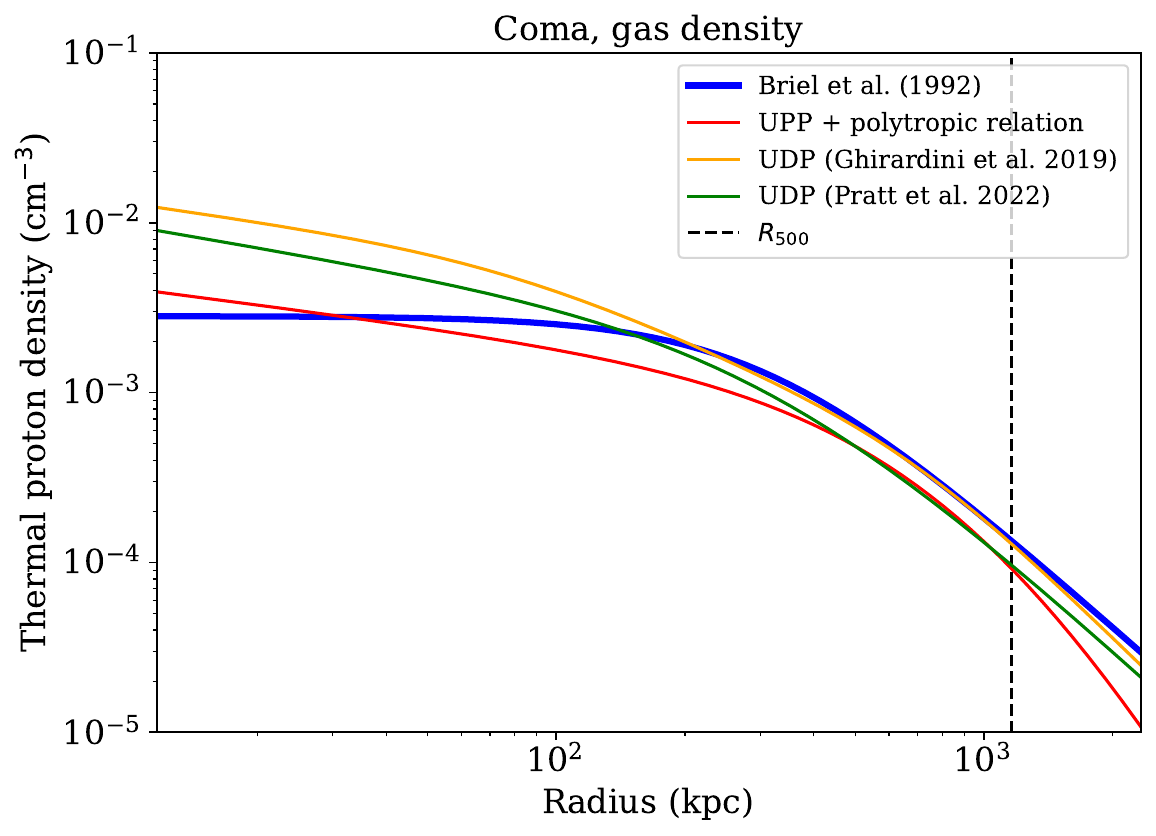}
\caption{Thermal proton density profiles for the Coma cluster: the blue line gives the best-fit model to the ROSAT data \citep[][]{Briel:1992xc}, the red line gives the density obtained from the \cite{Ghirardini_2019b} polytropic relation combined with the \cite{2010A&A...517A..92A} UPP profile, the green and orange lines give the UDP from \cite{Pratt:2022nrs} and \cite{Ghirardini_2019}, respectively.}
\label{fig: density}
\end{figure}

With the gas density in hand, we can derive the electron-, proton- and helium-density profiles by scaling through the mean molecular weights $\mu_{\rm gas} = 0.61$, $\mu_e = 1.16$, $\mu_{\rm p} = 1.39 $ and $\mu_{\rm He} = 14.6$ \citep[see, e.g.,][]{Adam:2020atc}. The proton density profile of the Coma cluster, as obtained from the best-fit model describing the ROSAT data \citep{Briel:1992xc}, is shown in Figure~\ref{fig: density}. It is compared to the model derived from our methodology, using the mass and redshift from the MCXC catalog \citep{Piffaretti2011}. The red line gives our reference model, i.e., the one obtained using the UPP profile combined with the polytropic relation. For further comparison, the UDP profile, as calibrated by \cite{Pratt:2022nrs} and \cite{Ghirardini_2019}, are given in green and orange, respectively. We can observe that the main differences between the data and the models, and among the models themselves, arise in the central part of the cluster. This reflects the increased intrinsic scatter among the cluster population relative to the self-similar approximation in the cluster cores, while the consistency significantly improves at $r \in [0.2 R_{500}, R_{500}]$ \citep[see, e.g.,][for details]{Ghirardini_2019}. More specifically in Figure~\ref{fig: density}, the Coma cluster is a merging system with a very flat core, thus presenting a smaller central density than that given in our mean model (we also refer to the appendix for further examples). The impact of the choice of the reference density model on our final results is discussed in Section~\ref{Sec 6: systematic studies}.

\subsection{Magnetic-field profile}

The profile of magnetic-field strength can be scaled to the thermal-gas density under several assumptions. Assuming the magnetic energy density to be proportional to the thermal energy, we have 
 \begin{equation}
     \langle B^2(r) \rangle = 2 \mu_0 P(r) / \beta_{\rm pl},
\label{eq:B_via_beta_plasma}
 \end{equation}
with $\mu_0$ the vacuum permeability. For the plasma, we set $\beta_{\rm pl} = 200$ following the results by \cite{Walker_2017} on the Perseus cluster. By combining the central magnetic field of the Coma cluster measured by \cite{Bonafede_2010} and the central pressure obtained by \cite{Planck2013X}, we would instead estimate $\beta_{\rm pl} = 77$. Alternatively, assuming that the magnetic field is frozen into the plasma and amplifies under pure adiabatic compression with magnetic flux conservation, we have 
 \begin{equation}
     \langle B^2(r) \rangle = B_{\rm ref}^2 \left(\frac{n_{\rm gas}(r)}{n_{\rm gas}(r_{\rm ref})}\right)^{4/3}.
\label{eq: B_via_compression}
 \end{equation}
The normalization $B_{\rm ref}$, taken at the radius $r_{\rm ref}$ is defined using the reference Coma cluster, for which detailed measurement are available in \cite{Bonafede_2010}.

\begin{figure}
\includegraphics[width=\columnwidth]{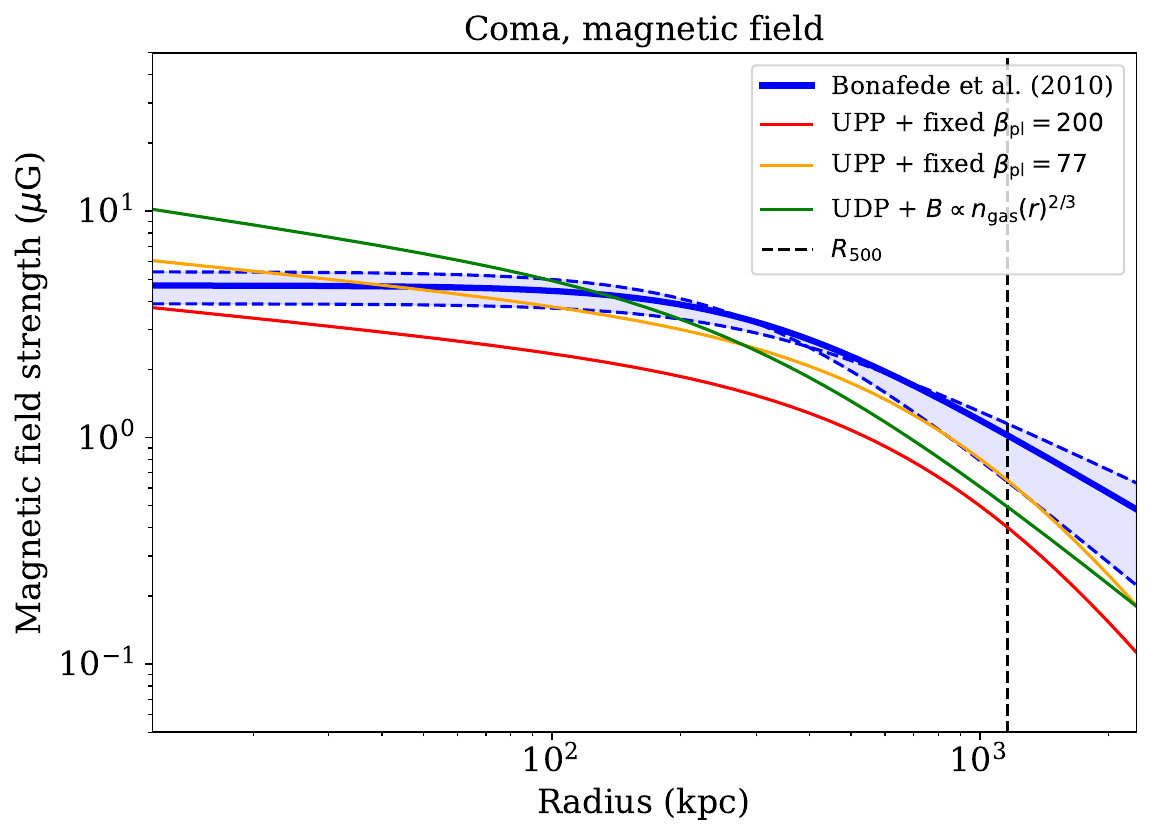}
\caption{Magnetic field strength profiles for Coma cluster: the blue line (and shaded region) show the best-fit model (and the constrained range) as obtained from Faraday rotation measure \citep{Bonafede_2010}, the red and orange lines give the model obtained when scaling the magnetic-field strength to the UPP and using $\beta_{\rm pl} = 200$ and $\beta_{\rm pl} = 77$, respectively, the orange line give our model when scaling the magnetic-field strength to the gas density, with the UDP from \cite{Pratt:2022nrs}.}
\label{fig: M_field}
\end{figure}

In Figure~\ref{fig: M_field}, we compare the magnetic-field profile of the Coma cluster estimated from Faraday rotation measures \citep{Bonafede_2010} to our models. The red line gives the profile estimated using Equation~\ref{eq:B_via_beta_plasma}, with $\beta_{\rm pl} = 200$, combined with the UPP from \cite{2010A&A...517A..92A}. The orange line is based instead on $\beta_{\rm pl} = 77$. The green line uses Equation~\ref{eq: B_via_compression} with the density estimated from the UDP calibrated by \cite{2009A&A...498..361P}. We observe that despite strong assumptions involved in our modeling, the prediction follows relatively well the measurement. This is also the case in the inner region of the cluster, where the environment is expected to play a major role for UHECR propagation.

%
 
In the following work, we use as a reference the UPP in Equation~\ref{eq:UPP} to derive the density through the polytropic relation in Equation~\ref{eq:ploytropic_DP}. The reference magnetic field is derived assuming constant magnetic to thermal energy density (Equation~\ref{eq:B_via_beta_plasma}) with $\beta_{\rm pl} = 200$. We compare the different assumptions in appendix for a set of clusters with different morphologies and discuss the impact of these assumptions on UHECR propagation in the following section.
\newpage
 \section{UHECR propagation in galaxy clusters} \label{Sec 4: Propagation}


\subsection{Interactions and diffusion in a cluster}
\label{sec:int_diff_cluster}
We compute the typical timescales for photo-hadronic and hadronic interactions of UHECRs in the cluster environment from a modified version of the Monte Carlo code \textit{SimProp} \citep[see][]{Aloisio:2012wj,Aloisio:2015sga,Aloisio:2016tqp}. We account for interactions with photons of the cosmic microwave and infrared backgrounds (CMB, CIB), as well as for hadronic interactions within the ICM.

Under the assumption of a monochromatic photon field of number density $n_{\gamma}$, the typical interaction rate  between a relativistic atomic nucleus ($A$) and a low energy photon is approximately $\tau_{A\gamma}^{-1} \simeq c \sigma_{A \gamma} n_{\gamma}$, where $\sigma_{A \gamma}$ represents the cross section of the process  and $c$ is the speed of light in vacuum. If a more realistic spectral energy distribution for the photon field is considered and the dependence of the cross section on the energy is taken into account, the interaction rate reads \citep{Aloisio_2013}:
\begin{equation}
\frac{dN_\text{int}}{dt} = \frac{c}{2\Gamma}\int_{\epsilon'_\text{th}}^\infty \sigma_{A\gamma}(\epsilon')\epsilon' \int_{\epsilon'/2\Gamma}^\infty \frac{n_\gamma (\epsilon)}{\epsilon^2} \,d\epsilon\,d\epsilon',
\end{equation}
where $\Gamma$ is the Lorentz factor of the interacting nucleus. Note that primed symbols (e.g.\ $\epsilon'$) refer to quantities in the nucleus rest frame, whereas unmarked symbols refer to quantities in the laboratory frame.

Though spallation processes between UHECRs and gas have negligible impact in the extragalactic medium, their role can be substantial in the ICM considering the effective time that relativistic  particles spend in this environment. The timescale for the spallation process reads:
\begin{equation}
\tau_{\rm spal} = {n_{\rm ICM} \, \sigma_{\rm sp} \, c }^{-1},
\label{eq:time_spal}
\end{equation}
where $n_{\rm ICM}$ is the  ICM gas density and $\sigma_{\rm sp}$ is the cross section for proton-proton or proton-nucleus interactions. 
This process has been implemented in \textit{SimProp} making use of the most recent hadronic model, {Sibyll 2.3d} \citep[][]{Riehn_2020}, a hadronic event generator. Details on the interface between the hadronic interaction model (HIM) and the in-source version of SimProp can be found in \cite{https://doi.org/10.48550/arxiv.2209.08593}.

In addition to interactions, diffusion in the magnetic field has to be taken into account. In fact, charged particles populating an astrophysical environment can be confined for a long time before escaping.  The diffusion timescale reads: $t_{\rm D}=R^2/D$,  where $R$ is the radius of the environment  , and where $D$ is the UHECR diffusion coefficient computed in the context of quasi-linear theory \citep{Lee_2017}. 
The expression of the diffusion coefficient is: $D \simeq c r_{L}^{2-\delta} \, l_{\rm c}^{\delta -1} /3$, where $r_L = E/qB$ is the particle Larmor radius, $l_{\rm c}$ is the coherence length of the magnetic field and $\delta$ is the slope of the turbulence power-spectrum, while $B$ is the strength of the turbulent magnetic field.  We assume $\delta = 5/3$ as prescribed for a Kolmogorov turbulence cascade.
 Following \cite{Subedi:2016xwd} and \cite{10.1093/mnras/stac1408}, we additionally consider the transition in the diffusion regime taking place when $r_L \gtrsim l_c$. 
In this energy range, the diffusion coefficient is estimated as $D = D_0 (r_L/l_c)^2$, where $D_0$ is the value of the diffusion coefficient computed at the energy $E_0$ such that $r_L(E_0)= l_c$. At the highest energies, the particle propagates ballistically so that the diffusion time tends to $R/c$.
\begin{figure}[t]
\centering
\includegraphics[width=\columnwidth]{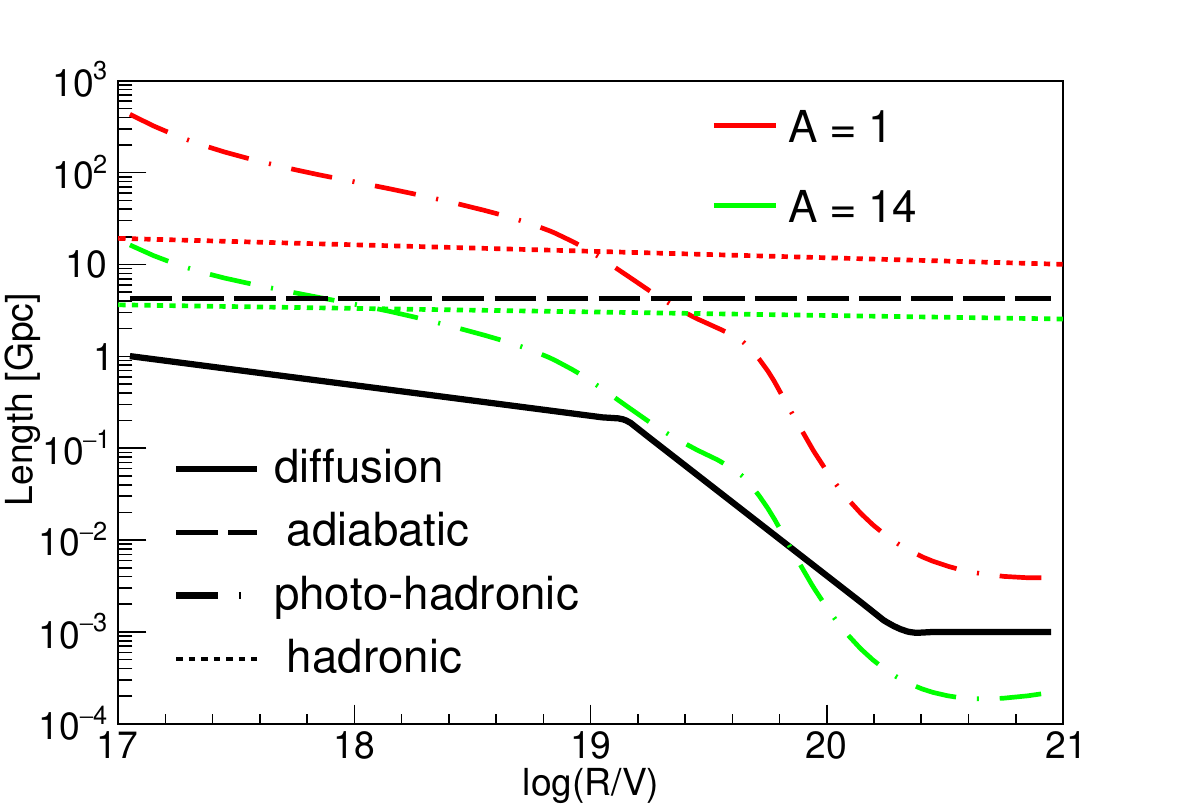}
\caption{Interaction and escaping lengths as a function of magnetic rigidity  at the center of a prototypical galaxy cluster: photo-hadronic interaction times (dashed-dot lines), spallation times (dashed lines) and diffusion times (solid lines) for protons (red) and nitrogen nuclei (green). The Hubble radius (corresponding to the age of the universe) is shown as a long-dashed line. Length scales have been calculated assuming the following parameters: $R_{500} = 1 \ \rm   Mpc$, $B = 1 \ \rm \mu G$,   $l_c = 10 \ \rm kpc$, $n_{\rm ICM} = 1 \cdot 10^{-4} \ \rm cm^{-3}$.}
\label{fig: timescale}
\end{figure}

Figure~\ref{fig: timescale} summarizes the typical length-scales for interactions and escape in the source environment for a prototype cluster (see caption). The interplay between length-scales governs the shape of the UHECR flux as well as the nuclear composition at the escape from the cluster.
The shortest length-scale for protons is always dictated by diffusion; this means that some protons can escape from the environment.
For nuclei (e.g.\ nitrogen in Figure \ref{fig: timescale}), photo-interaction lengths are the shortest at high rigidities for the chosen parameters of the cluster (see caption). Clusters with larger magnetic fields also present higher target densities, which reduces the hadronic interaction length and makes hadronic interactions predominant at lower rigidities. 
\newpage
\subsection{Implementation of ICM propagation in SimProp}\label{sec: implementation}

In order to model the UHECR transport in clusters, we have developed an extension of {\it SimProp}.
This software has been used so far in the context of the extragalactic propagation of UHECRs \citep[see for instance][]{combFit, Abdul_Halim_2023,Luce_2022}. SimProp implements different photo-disintegration cross sections and different models for the CIB. In this work, we adopt TALYS \citep{TALYS_1,TALYS_2} for the photo-disintegration cross sections and the CIB model of \cite{Gilmore_EBL}, which are both representative of the state of the art. SimProp is a monodimensional propagator. Assuming spherical symmetry, all the particles are propagated along an axis of the cluster until they reach $3 \times R_{500}$, a distance beyond which the ICM has negligible impact with respect to the extragalactic medium.

We also consider the impact of the magnetic field on UHECR propagation. { Charged particle moving through a uniform magnetic field undergo an angular  deflection upon traversing a distance, $l_c$, of $ \simeq \dfrac{l_{c}}{r_L}$. A particle of energy $E$ and charge $q=Ze$ traversing a distance $L$ suffers an overall angular deflection given by $\theta (E,Z) \simeq \bigg(\dfrac{L}{l_{c}}\bigg) \cdot  \theta$ \citep{Hooper_2007}, which depends on the properties of the environment ($B$, $L$ and $l_c$) and  of the particles ($E$, $Z$). Such deflections result in an increase in the effective propagation length, $L_{\mathrm{eff}}$,} in the ICM given by \citep{PhysRevD.72.043009}:
\begin{equation}
    \dfrac{L_{\mathrm{eff}}}{L}
    \simeq 65\, \bigg( \dfrac{E/Z}{10^{20} \, \rm eV/26} \bigg)^{-2}  \bigg( \dfrac{L}{1 \ \rm Mpc}\bigg) \bigg( \dfrac{l_{c}}{10 \ \rm kpc} \bigg) \bigg( \dfrac{B}{1 \ \rm \mu G} \bigg)^2
\end{equation}
Knowing the properties of the cluster, it is possible to compute the effective length and therefore the effective time that a particle spends in the environment.

The propagation inside the cluster environment is determined according to the following methodology: 
1) The propagation axis is divided in a given number of steps, $n_{\rm steps} \geq A$, with $A$ atomic mass of the injected nuclei, sufficiently large to sample the interactions;
2) UHECRs are injected at a given point in the cluster and the propagation is performed only along the chosen axis; 3) The typical length-scales are dependent on the position, according to the magnetic-field and gas density profiles. The probability of interaction or escape changes as a function of the radius; 4) Particles are moved to the following step if the interaction probability is smaller than the escape one, otherwise they lose energy and their byproducts are accounted for in the following steps of the propagation; 5) Once a particle has reached the border, if the diffusion probability is larger than the interaction one, this particle escapes from the cluster environment and is propagated through the extragalactic medium; 6) Particles that spend a time greater than the age of the universe in the environment are considered trapped and are not propagated anymore. This is a conservative assumption: the dominant time is the minimum between the age of the cluster and the age of the oldest accelerator inside it, both smaller than the age of the universe.

\section{UHECR flux escaping the ICM}\label{Sec: 5-Param}
\begin{figure}[t]
\centering
\includegraphics[width=\columnwidth]{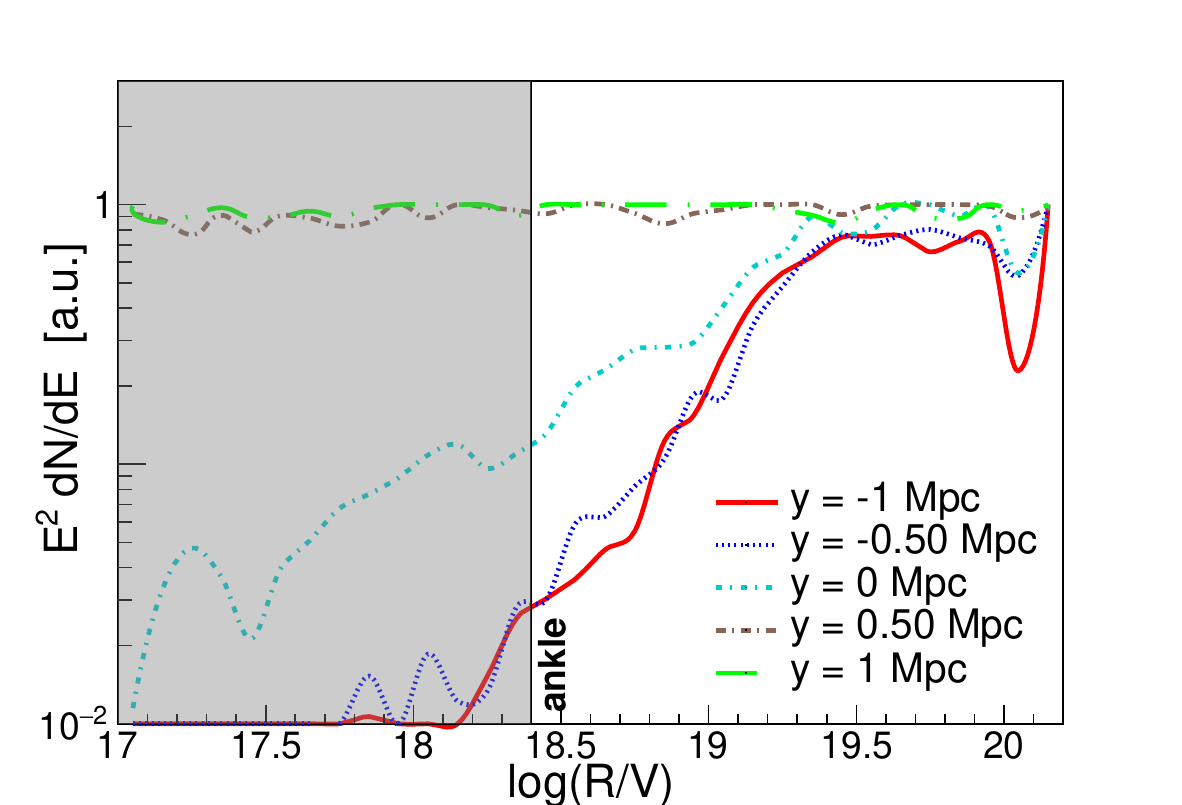}
\caption{Escaping proton spectra from a cluster of $M = 10^{14} \ M_{\odot}$ as a function of the injection point. Positive (negative) $y$ valued correspond to positions closer to the nearest (furthest) edges of the cluster. The spectra are normalized to that expected without interactions in the ICM. The vertical line shows the ankle energy.}
\label{fig: inj_point}
\end{figure}

Once particles escape from the magnetized environment, it is possible to evaluate what is the impact of the ICM on the UHECR spectrum as a function of the injection point. We inject $10^4$ particles logarithmically distributed in the energy range $10^{17} - 10^{21} \ \rm eV$. The results are shown in Figure~\ref{fig: inj_point}, where the escaping fluxes are represented as a function of rigidity. The spectra are normalized to the spectrum expected if interactions and diffusion in the ICM were neglected. One can notice how the closer the injection is to the nearest edge of the environment (at $\approx +1$\,Mpc), the more the escaping flux coincides with the injection spectrum: the gas and magnetic field densities are low and the propagating particles are less affected. If instead the UHECRs cross the center of the cluster ($y \leq 0$), the flux is reduced at low energies due to the trapping by the magnetic field \citep[the so-called magnetic horizon for extragalactic propagation, e.g.][]{PhysRevD.71.083007, Gonz_lez_2021}. At the highest energies, fluctuations in the ICM transmission are an artifact of the normalization procedure, in a regime where interactions of UHECRs with the CMB are important. 

\begin{figure}
\centering
\includegraphics[width=\columnwidth]{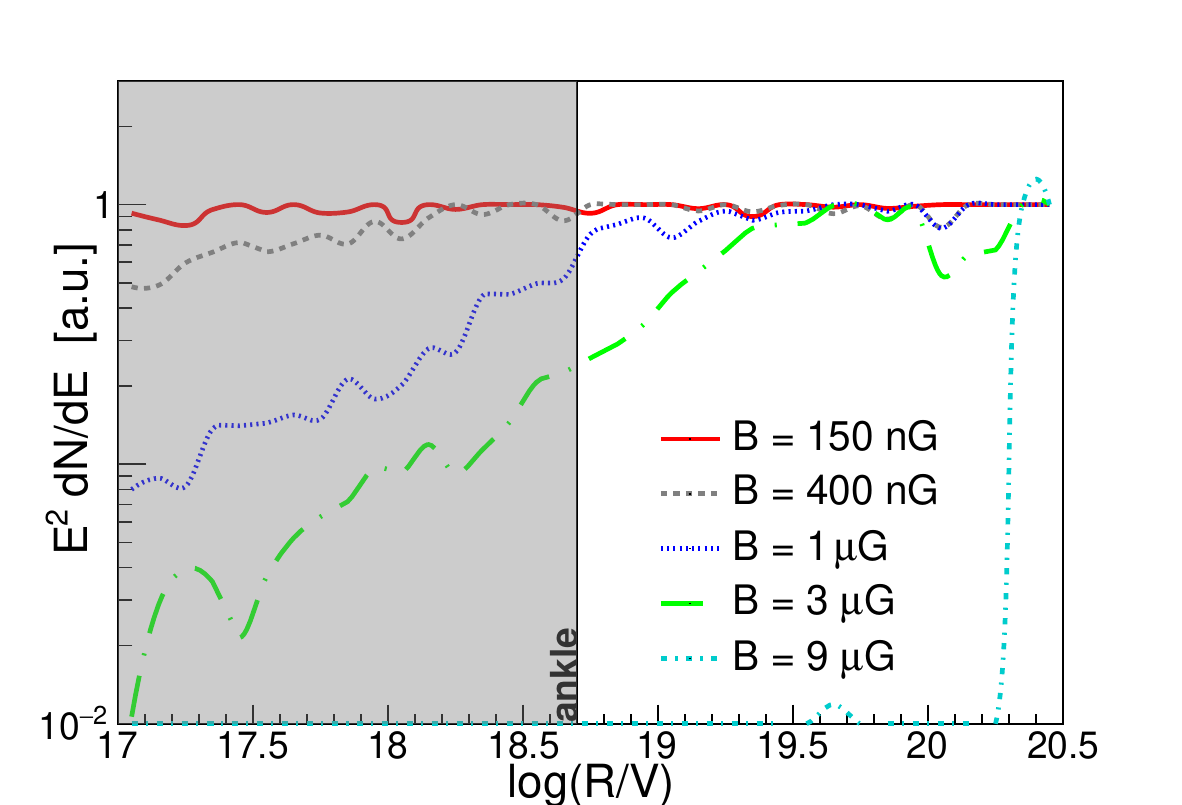}
\caption{Escaping proton spectra from cluster of different magnetic field values at the center, taken at 1 kpc (see legend), assuming injection at the center of the cluster. The spectra are normalized to that expected without interactions in the ICM. The vertical line shows the ankle energy.}
\label{fig: mass}
\end{figure}

More massive clusters present more intense magnetic field at the center of the cluster, which shortens the magnetic horizon of UHECRs. 
The impact of the cluster magnetic field on the propagation of UHECRs  is illustrated in Figure~\ref{fig: mass}, where the escaping fluxes are shown as a function of rigidity, in this case assuming only protons at the injection.

\subsection{Parametrization of the UHECR escaping  flux}
\label{sec:param}

We provide a parametrization of the escaping fluxes as a function of the mass of the cluster $M$, of the position of injection point $y$ and of the nature of the accelerated particles (protons or nuclei) in order to describe the escaping flux above the ankle. Four representative nuclear masses are studied: $^{1}$H, $^4$He, $^{14}$N, $^{28}$Si. The contribution from iron nuclei is neglected, as few, if any, are expected from simple cosmological models that describe data from the Pierre Auger Observatory \citep{combFit,Abdul_Halim_2023,Luce_2022}.

We notice in Figure~\ref{fig: inj_point} that a cluster mostly affects the escaping spectrum when UHECRs cross its center. In fact, it is the place where the magnetic field is most intense and where the target density is the highest. For this reason, sources placed at $y \leq 0$ would have an escaping fluxes shaped by the propagation in the cluster environment, while the effect is weaker for host sources placed at $y > 0$ where the traversed magnetic field is milder.
For this reason, we assume clusters to be transparent for accelerators placed at $y > 0$, while we provide a single parametrization of the transparency of the clusters for $y \leq 0$. We define the transparency $f(R)$ of a given cluster as the escaping flux divided by the one expected without interactions in the ICM. We approximate the transparency  as a function of rigidity $R$ by a broken power-law, with full transparency at the highest energies:
\begin{equation}
\label{eq:param}
\log \ f(R) = \begin{cases}
 \Gamma \ \log(R/\rho) & R \leq \rho, \\
{0} & R \geq \rho.
\end{cases}
\end{equation}
We notice that, in our equation above, the break rigidity, $\rho$, depends on the mass of the cluster, $M$, following to first order:
\begin{equation}
    \log{\rho} = \log \rho_0 +\xi\log(M/10^{15} M_{\odot}).
\end{equation}

We parametrize the low-rigidity slope, $\Gamma$, of the transparency function so that it reaches a maximum value of $2$ at high cluster masses and softens at lower masses:

\begin{equation}
    \Gamma = \dfrac{2}{1 + \bigg(\dfrac{M}{M_{\rm free}}\bigg)^{-\sigma}}.
\end{equation}

We find that $\log(M_{\rm free}/M_\odot)  = 14.4 \pm 0.5$ is consistent with the transparency functions of  both nuclei and protons. The parameter $\sigma$ governs the evolution of the index with cluster mass. Also in this case, we find a common value $\sigma = 0.25 \pm 0.10 $ for both nuclei and protons.

\begin{figure}
\centering
\includegraphics[width=\columnwidth]{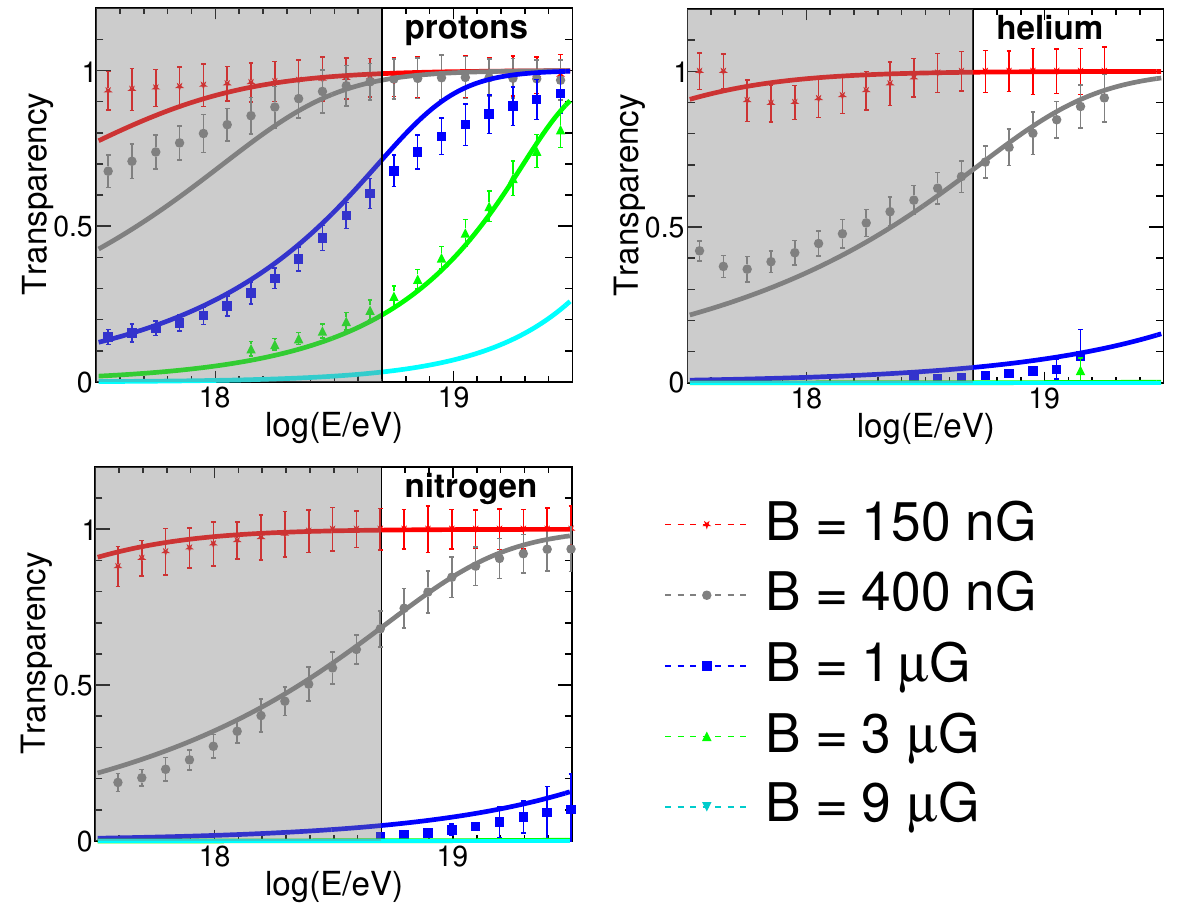}
\caption{Transparency as a function of energy for protons, helium and nitrogen nuclei for different cluster magnetic field (see legend), assuming an an injection point at the center of the environment. The points show the results obtained from the simulations with errors resulting from the number of injected particles. The solid lines display the proposed parametrization. The vertical line shows the ankle energy.}
\label{fig: Param_protons}
\end{figure}

The parameters are determined by fitting the model in Equation~\ref{eq:param} to the escaping fluxes for different position of the sources at $y \leq 0$ and for different cluster masses, considering either protons only or nuclei only. We find best parameter values of $\log(\rho_0/\rm V) = 20.0 \pm 0.2$ for protons and  $\log(\rho_0/\rm V) = 24.3 \pm 0.3$ for nuclei, while $\xi = 0.6 \pm 0.1$ for protons and  $\xi = 1.7 \pm 0.2$ for nuclei. The comparison to simulated data is shown in Figure~\ref{fig: Param_protons}, for an  injection at the center of the cluster.

The two parameters that influence the rigidity at which the transition happens, i.e.\ $\rho_0$ and $\xi$, are larger for nuclei than for protons. This is due to the fact that the nuclei interact more than protons in the ICM, as shown in Figure~\ref{fig: timescale}; for this reason the transition to $f(R) = 1$ happens at higher rigidities for nuclei.

%

Our simulations show that clusters of mass $M = 10^{14} M_\odot$ or $M = 10^{15} M_\odot$, with central magnetic fields of $3$ and $9\,\mu$G respectively, are able to trap nearly all protons up to the ankle. For lower magnetic fields, the effect of the ICM on protons is quite negligible above the ankle.
Similar conclusions can be drawn for nuclei; nonetheless it is important to stress that they are fully disintegrated up to at least the ankle for clusters with central magnetic fields larger than $1\,\mu$G.

The proposed parametrization describes well the impact of the galaxy cluster on the escaping flux above the ankle. The approximation of considering the environment as transparent for sources at $y > 0$ describes well the results of the simulation for weakly magnetized clusters. For clusters with $B \geq 3 \ \mu$G, the proposed parametrization for $y > 0$ overestimates the escaping fluxes on average by 0.4\,dex.

Overall, we can conclude that only a few percent of nuclei can escape from clusters with $B \geq 1\,\mu$G
up to energies of $10^{19}$\,eV for He and $10^{19.5}$\,eV for N. Protons are strongly suppressed as well in the most massive clusters: only 40\% escape at $10^{19} \ \rm eV$ for a central magnetic field of $3\,\mu$G while practically none escape at this energy for $B \geq 9\,\mu$G.
Galaxy clusters are thus hostile environments for UHECRs. The filtering is more intense for nuclei, which are fully disintegrated in the most massive clusters even in the outer regions of the environment. 

\subsection{Impact of our assumptions}\label{Sec 6: systematic studies}

In this investigation, many assumptions have been made. This section aims to discuss their impact.

{The most impacting assumption is the parametrization of the diffusion time, which is based in this work on the scaling laws expected from diffusion theory (see Sec.~\ref{sec:int_diff_cluster}).
Using instead the equation~20 of \cite{Harari:2013pea}, which is based on Monte-Carlo simulations, results in an even more opaque environment, with a transparency reduced by a factor $\simeq 2$ above the ankle 
for protons injected in the center of the cluster.}

{The coherence length of the magnetic field in the ICM also influences our results. In this work, all the clusters are assumed to have the same coherence length, $7$\,kpc, based on observations of Coma. More detailed constraints on this quantity would be instrumental in determining the UHECR transparency of clusters on a case-by-case basis.}

Another important topic to be discussed is the assumption of the magnetic-field and gas density profiles. Instead of the reference model detailed in Section~\ref{Sec: 2-Galaxy cluster}, we performed the same analysis as in Section~\ref{sec:param} using the best-fit models shown as blue lines in Figure~\ref{fig: density} and \ref{fig: M_field} (see also appendix) for three different clusters: Virgo \citep{2016A&A...596A.101P}, Coma \citep{2021A&A...648A..60A} and Perseus \citep{Churazov_2004}. In the three examined clusters, the differences in transparency are at maximum of the order of $1\%$, irrespective of their morphology, therefore it is possible to affirm that the assumption of a UPP does not influence the main results of this work. The self-similar framework is largely driven by the cluster mass, which can be a difficult quantity to measure. The accuracy of cluster-mass estimates is thus expected to be a primary source of uncertainty. One should nonetheless note that, under our reference approach, the magnetic-field strength scales to first order as $B \propto \sqrt{P_{500}} \propto M_{500}^{1/3}$, so that only an order-of-magnitude uncertainty has a strong impact on the UHECR transparency illustrated in Figure~\ref{fig: Param_protons}.

We only considered interactions with CMB and CIB photons, neglecting the contribution of stellar and dust-grain emission in the cluster. \cite{Harari_2016} estimated such galactic contributions to be comparable to the CIB.  Neglecting the galactic emission does not affect our results. This can be understood looking at Figure~\ref{fig: timescale}, where the change of slope at around $10^{19.6} \ \rm eV$ in the curves labeled ``photo-interaction'' corresponds to  the transition from lower-energy interactions with CIB to higher-energy interactions with CMB. For protons, the CMB is the only relevant field within the Hubble time. For nuclei, we investigated the impact of doubling the CIB density to model the galactic emission. The transparency changes only by few percents. For this reason, as in \cite{Harari_2016, Hussain2021}, the galactic emission is neglected.

In this work, we adopt Sibyll2.3d \citep{Riehn_2020} as HIM. The systematics related to the use of this specific HIM cannot be explored currently, because no other HIM are currently implemented in SimProp. This investigation should be discussed in  future works. 

\section{Discussion}
\label{Sec: 9 Conclusion}
In this work, we develop a detailed model to explore the extent to which galaxy clusters impact UHECR propagation. In particular, the modelling of the cluster environment and the use of a HIM for propagation in this environment represent novelties for UHECR propagation studies.

We work under the assumption of self-similarity. From this assumption, it is possible to derive the important quantities for UHECR propagation, namely the magnetic-field and the gas density profiles given the mass and the redshift of the clusters. We find that the cluster environment acts as a high-pass filter, allowing a fraction of UHE protons to escape while the UHE nuclei interact with the gas and photons present in the ICM.

This work presents some advances with respect to the previous literature. The use of a software dedicated to the treatment of the cluster environment is new in UHECR physics; the conclusions of this work are in line with other works which predicted that galaxy cluster are hostile environment for UHE nuclei \citep{Kotera_2009, Harari_2016, Fang_2018}, while they have a weaker although non-negligible effect on the propagation of UHE protons.
For example, \cite{Harari_2016} suggest in their Figure~6 that, for a cluster with a central magnetic field of $B = 1 \ \mu G$ and coherence length of $l_{\rm c} = 10 \ \rm kpc$, the environment is completely transparent for protons above the ankle energy ($ > 5 \cdot 10^{18} \ \rm eV$), while it affects slightly the escaping flux of intermediate-mass nuclei (70\% transparency for carbon nuclei at 10\,EeV) and heavy nuclei (9\% for iron nuclei at 10\,EeV). In our case, neglecting the differences in the profiles {and parametrization of the diffusion coefficient}, we can compare these results with those obtained for a structure of mass $M = 10^{13} \ M_{\odot}$, which corresponds to a central magnetic field  of ${\simeq}\, 1 \, \mu G$. The two results are {comparable} for protons, while our model predicts a significantly larger depletion of intermediate nuclei: 10\% transparency in our work instead of 70\% in \cite{Harari_2016}. We confirmed through simulations that this difference arises from the different treatment of the hadronic interactions; the use of a HIM, instead of a simple analytical model, can strongly influence the propagation of the cascade, affecting both the fraction of energy lost and the fragmentation of heavy nuclei. No direct comparison can be performed with the work of \cite{Fang_2018}, which shows the cumulative spectra of UHECRs at Earth that escaped from a population of sources, nor with that of  \cite{Kotera_2009}, where the UHECR flux escaping from a single cluster is arbitrarily normalized.

The present work leads to important conclusions for the emerging field of UHECRs astronomy. Two different trends can be observed in the mass composition of UHECRs measured with the Pierre Auger Observatory \citep{Composition_2019}: a transition from heavy to light mass composition is observed up to $10^{18.3} \, \rm eV$, while data at higher energies suggest a transition to intermediate-heavy masses. Based on our simulations, we should not observe UHE nuclei coming {from the inner regions} of massive galaxy clusters above the ankle energy. This includes in particular the Virgo cluster, the closest galaxy cluster to us \cite[$d \simeq  16 \ \rm Mpc$, $M \simeq 1.2 \cdot 10^{14} \ M_{\odot}$, {from}][] {2016A&A...596A.101P}. The Pierre Auger Collaboration indeed does not see any indication of excess in this direction, which could have an important implication as pointed in \cite{https://doi.org/10.48550/arxiv.2108.10775}. Assuming that the UHECR production rate follows the star-formation rate or stellar mass of nearly half million of galaxies, \cite{Biteau_2021} found that the computed skymaps should show some excess in the direction of the Virgo cluster, not present in the observed skymaps \citep{Abreu_2022}. Our work confirms that this tension is lifted by magnetic trapping of UHECRs in Virgo, as was already  hypothesized in \cite{https://doi.org/10.48550/arxiv.2108.10775} through a more naive argument (confinement time  greater than the ballistic one). The result of our work reduces the discrepancies between the arrival direction model and the data, justifying the lack of UHE nuclei in the directions of the galaxy clusters and thus suggesting interesting pathways to investigate composition anisotropies. 

Another application of our work is related to the dipole observed by the Pierre Auger Observatory above 8\,EeV, whose direction is qualitatively explained from the distribution of local extragalactic matter and UHECR deflections in the Galactic magnetic field \citep{doi:10.1126/science.aan4338}. The strong contribution to the dipole from the Virgo cluster inferred e.g.\ by \cite{Ding_2021} {assuming that the UHECR production rate follows the distribution of matter} should be significantly lowered when accounting for magnetic trapping and shadowing in the ICM. This statement is true also for the Perseus cluster \citep[$d \simeq  74 \ \rm Mpc$, $M \simeq 5.8 \cdot 10^{14} \ M_{\odot}$, {from}][]{Urban:2013xca}, in the direction of which the Telescope Array collaboration claims an indication of excess above $5.7 \cdot 10^{19} \ \rm eV$ \citep{2021arXiv211014827T}. From the analysis, it cannot be excluded that the UHECRs come from the {vicinity or outer shocked region of the cluster}. This work tends to exclude the possibility that Telescope Array and the Pierre Auger Observatory see UHE nuclei accelerated by a host source close to the center of the Perseus or Virgo cluster; either they have to come from the {cluster outskirts} or they have to be UHE protons, both primaries or secondaries due to the fragmentation of heavy nuclei in the environment surrounding the accelerator.

An interesting result of this work concerns the role of filaments of the cosmic web \citep{2008PhRvD..77l3003K}.
In these regions \citep{2022MNRAS.512..945C}, the turbulent component  (inferred to be at the $\simeq \rm nG$ level) is weaker than the regular magnetic field ($\simeq 30\, \rm nG$), both much weaker than central magnetic fields of clusters. This means that while UHECRs are trapped in the central regions of galaxy clusters,  they can escape from filaments as stated in \cite{2019SciA....5.8227K}. If, as suggested by the authors, UHECRs are correlated with filaments connected to the Virgo cluster, they should escape from galaxies in the filaments.

A possible critical aspect beyond the scope of this work concerns the secondary production. In fact, interactions of UHECRs lead to an excess of secondaries, namely secondary cosmic rays, neutrinos and photons, which can escape from the environment and could be in tension with the current measurements. It should be noted that secondary protons produced by the fragmentation of heavy nuclei would remain trapped in the environment, so that they would not show up at lower energies. A natural step forward in this analysis would concern the multi-messenger connection, by taking into account the emission and propagation of photons and neutrinos in the environment. In this way, it would be possible to compare the escaping gamma rays with the possible excess observed by \textit{Fermi}-LAT in the direction of the Coma cluster \citep[$d \simeq  100$ Mpc, $M \simeq 7 \cdot 10^{14}$ M$_{\odot}$, {from}][]{Planck2013X}, as well as to determine the expected sensitivity of upcoming gamma-ray and neutrino facilities at higher energies. 

\begin{acknowledgments}
\section*{Acknowledgments}
{The authors would like to thank the reviewer for constructive suggestions, which helped to improve the quality of this work.} AC and JB gratefully acknowledge funding from ANR via the grant MultI-messenger probe of Cosmic Ray Origins (MICRO), ANR-20-CE92-0052. AC acknowledges co-developers of SimProp  for useful feedback in the development of the software.
\end{acknowledgments}

\newpage

\appendix
In this appendix, we show how the models of the gas-density and magnetic-field profiles discussed in Section~\ref{Sec: 2-Galaxy cluster} match the available measurements. We use reference clusters for which magnetic field profile estimates are available in the literature \citep{Vacca}, using the mass and redshift taken from the MCXC catalog \citep{Piffaretti2011}. For a sufficiently large sampling of mass and dynamical state, we select the Perseus, Coma, A\,194, Hydra\,A, A\,2634 clusters. For each of them, a best-fit model (or a model that was matched to the data, in the case of the magnetic field) is reported: \cite{Churazov_2004,2006MNRAS.368.1500T, Walker_2017} for Perseus, \cite{Briel:1992xc} for Coma, \cite{Govoni:2017qmd} for A\,194, \cite{Vogt_2003,En_lin_2006,Laing_2008,Kuchar:2009jj} for Hydra\,A and \cite{1997A&A...327...37S} for A\,2634. These are compared to experimental data from the Plank/ROSAT project \citep{2012A&A...541A..57E},
the XCOP project \citep{2017AN....338..293E}, 
and the ACCEPT data base \citep{2009ApJS..182...12C} 
whenever available and to the models outlined in Section~\ref{Sec: 2-Galaxy cluster}. The red lines correspond to the reference model of this work: UPP and polytropic assumption for determining the thermal gas density, constant magnetic to thermal energy density for magnetic field. The other possible choices are detailed in Section~\ref{Sec: 2-Galaxy cluster}.

\begin{figure}[h!]
\label{fig:cluster_profiles}
\figurenum{7.1}
\plotone{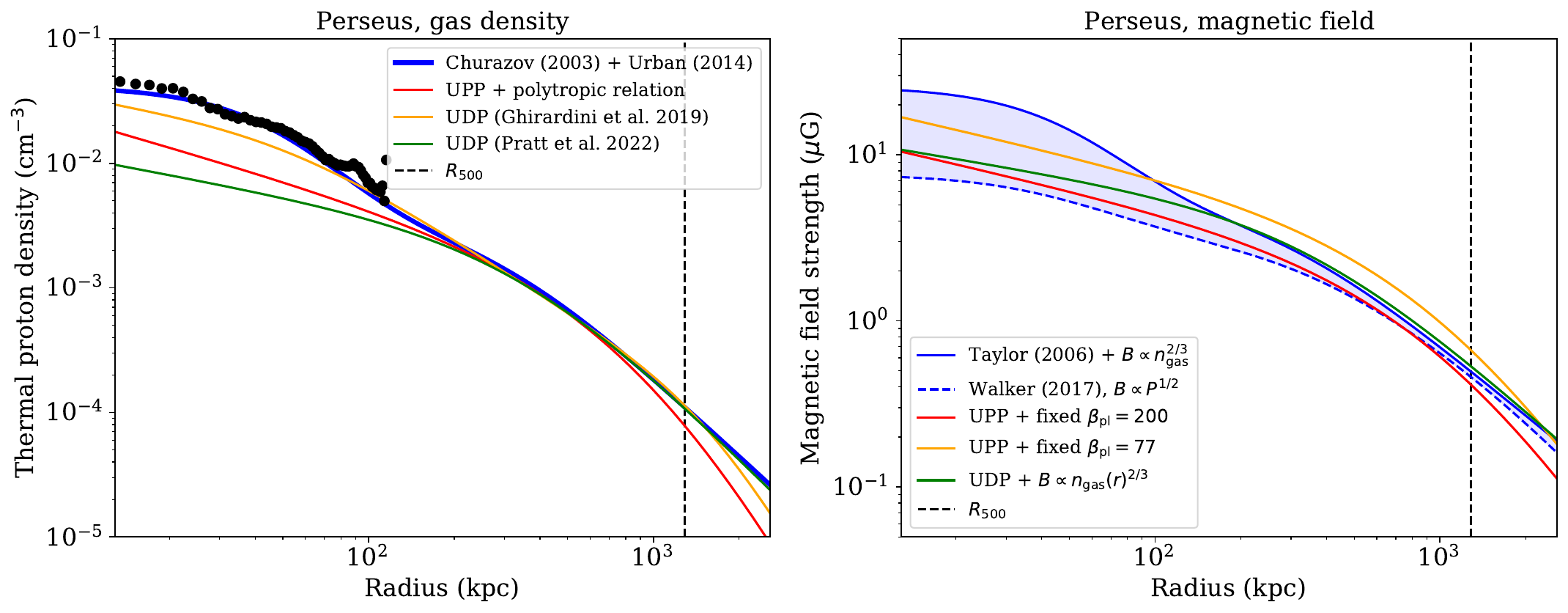}
\caption{Perseus cluster ($M \simeq 5.8 \cdot 10^{14} \ M_{\odot}$) : typical relaxed cluster with a dense core. {The complete figure set (11 images) is available in the online journal.}}
\end{figure}

In Figure~\ref{fig:cluster_profiles}, the best-fit models of the thermal-gas density profiles are matched to X-ray data. These models are beta models (or the sum of beta models), which present a flat core by construction and thus may be over-simplistic. Magnetic-field estimates derive from the matching of cluster simulations to radio data and/or Faraday-rotation measure, both generally assuming a scaling between the magnetic-field strength and the thermal-gas density. The uncertainties in these estimations are often not well defined. Some of the magnetic-field profiles from the literature may not be fit to the data. We conclude that our reference model is in acceptable agreement with measurements from the literature in the explored mass and redshift range, given the above-mentioned caveats in the measurement uncertainties and simplifying modeling assumptions.

\newpage

\figsetstart
\figsetnum{7}
\figsettitle{Thermal-proton and magnetic-field densities of reference clusters.}

\figsetgrpstart
\figsetgrpnum{7.1}
\figsetgrptitle{Thermal-proton and magnetic-field densities of Perseus}
\figsetplot{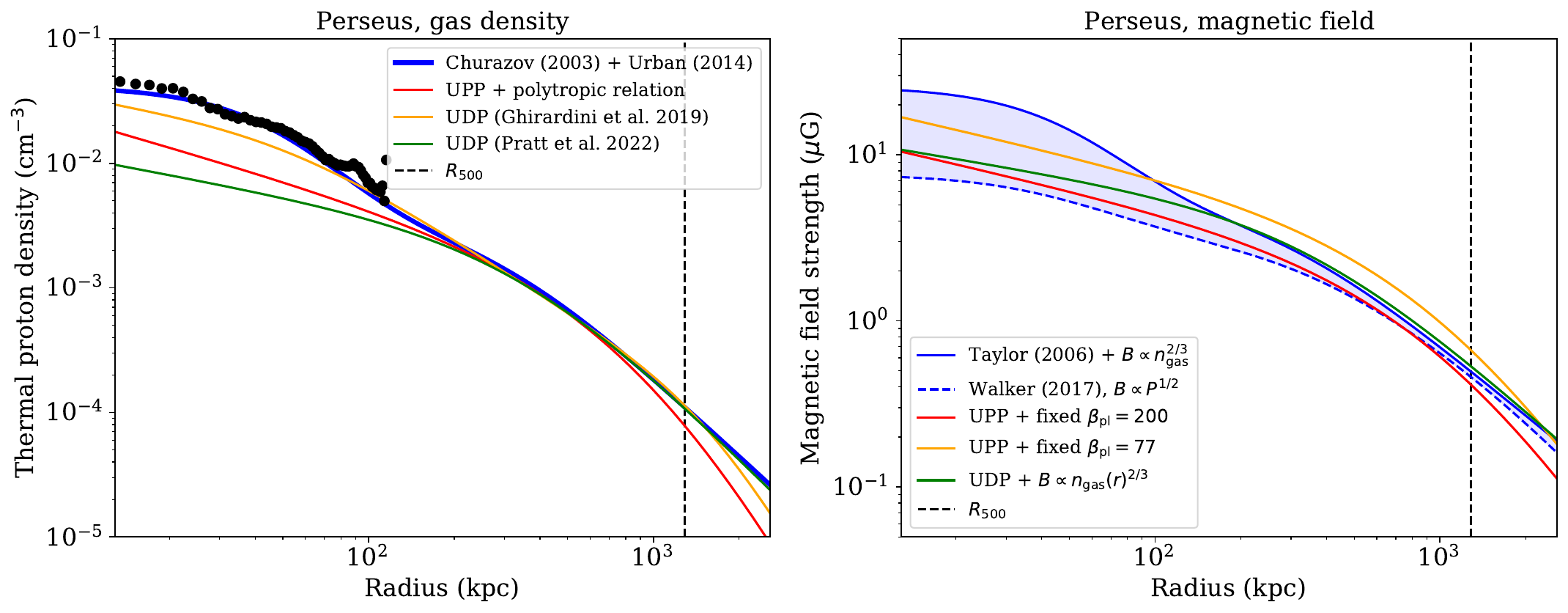}
\figsetgrpnote{Perseus cluster ($M \simeq 5.8 \cdot 10^{14} \ M_{\odot}$) : typical relaxed cluster with a dense core. }
\figsetgrpend

\figsetgrpstart
\figsetgrpnum{7.2}
\figsetgrptitle{Thermal-proton and magnetic-field densities of Coma}
\figsetplot{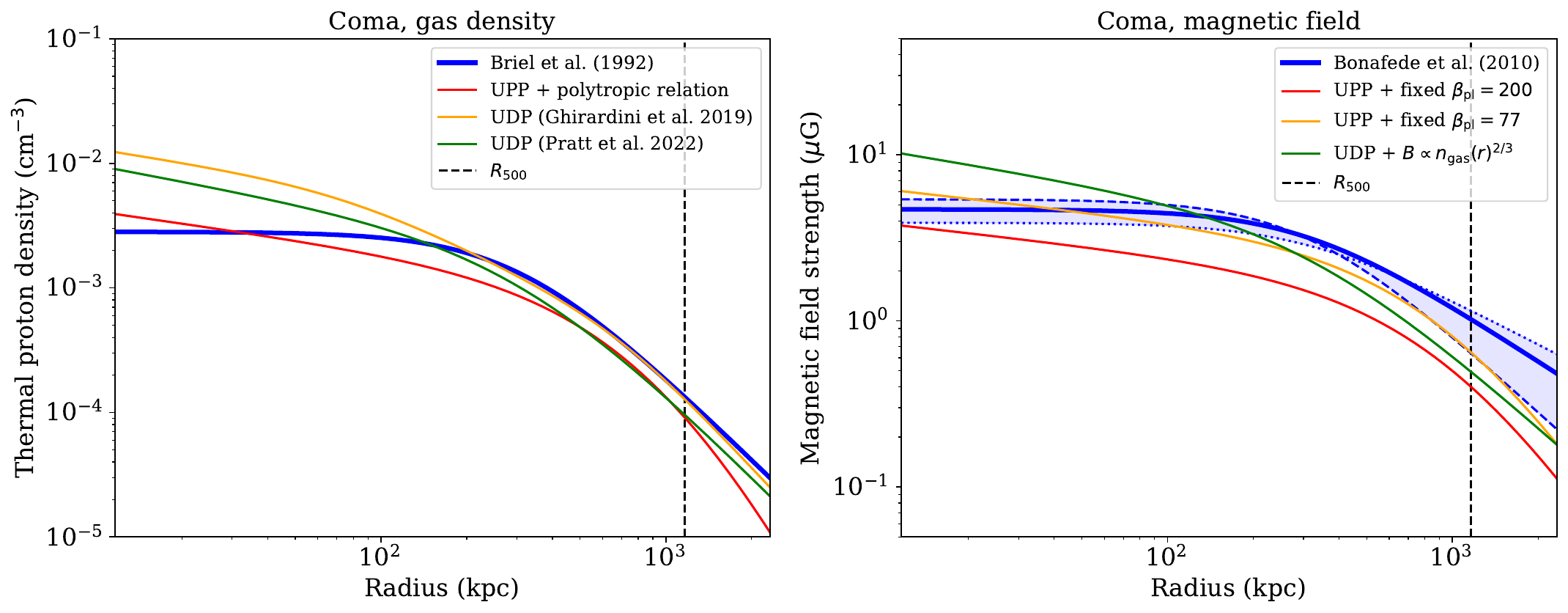}
\figsetgrpnote{Coma cluster ($M \simeq 7 \cdot 10^{14} \ M_{\odot}$) : merging cluster with flatter core.}
\figsetgrpend

\figsetgrpstart
\figsetgrpnum{7.3}
\figsetgrptitle{Thermal-proton and magnetic-field densities of A194}
\figsetplot{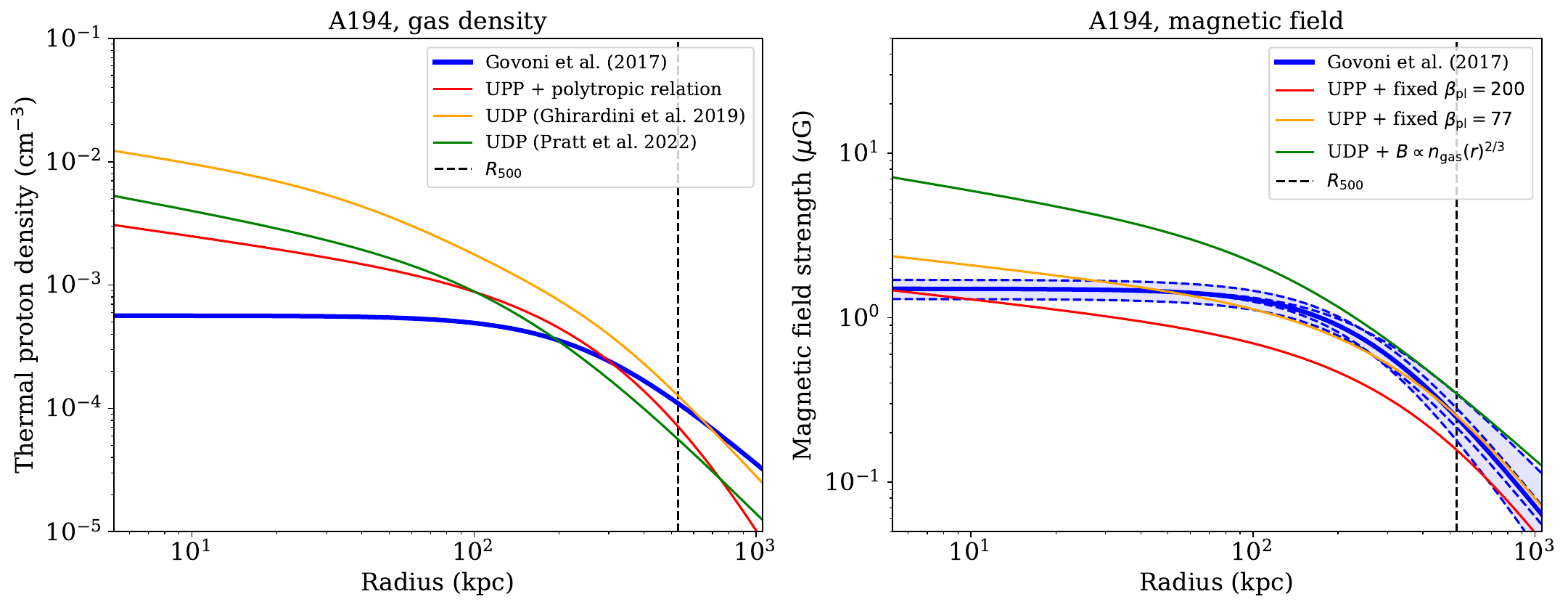}
\figsetgrpnote{A\,194 cluster ($M \simeq 4 \cdot 10^{13} \ M_{\odot}$): merging cluster with flatter core.}
\figsetgrpend

\figsetgrpstart
\figsetgrpnum{7.4}
\figsetgrptitle{Thermal-proton and magnetic-field densities of Hydra}
\figsetplot{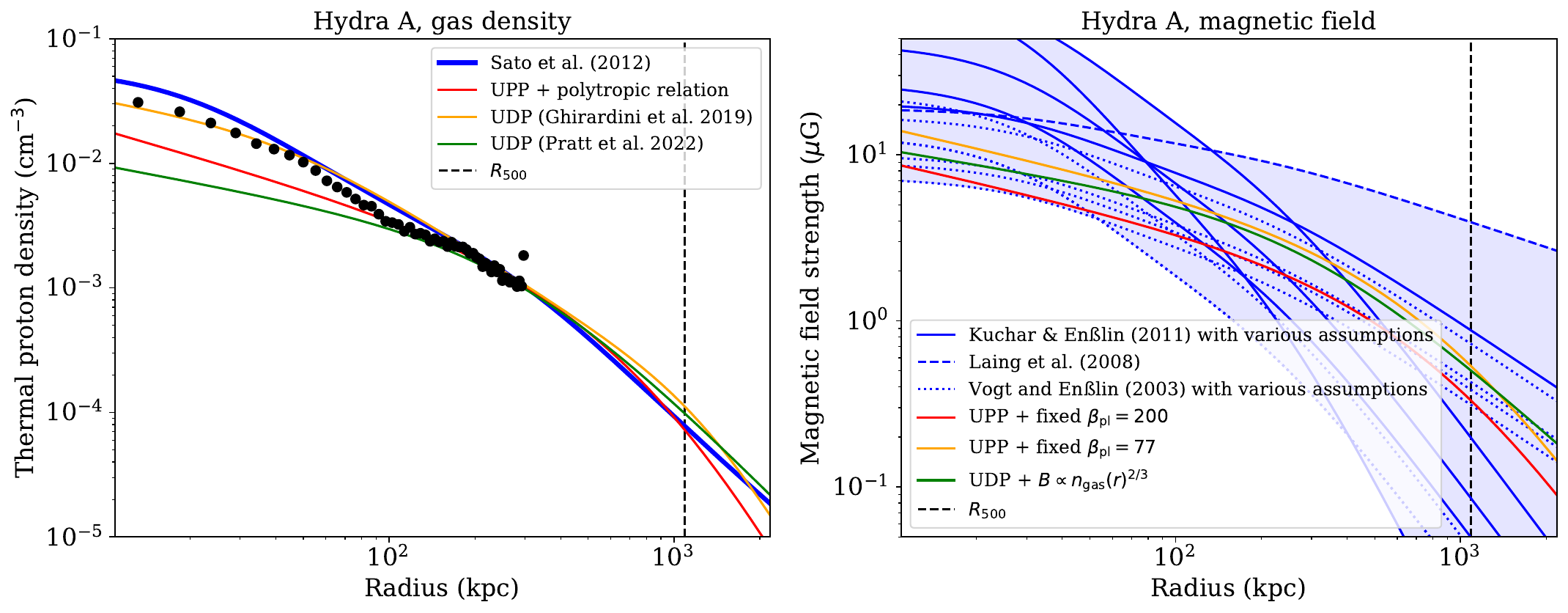}
\figsetgrpnote{Hydra\,A cluster ($M \simeq 3 \cdot 10^{14} \ M_{\odot}$) : typical relaxed cluster with a dense core.}
\figsetgrpend

\figsetgrpstart
\figsetgrpnum{7.5}
\figsetgrptitle{Thermal-proton and magnetic-field densities of A2634}
\figsetplot{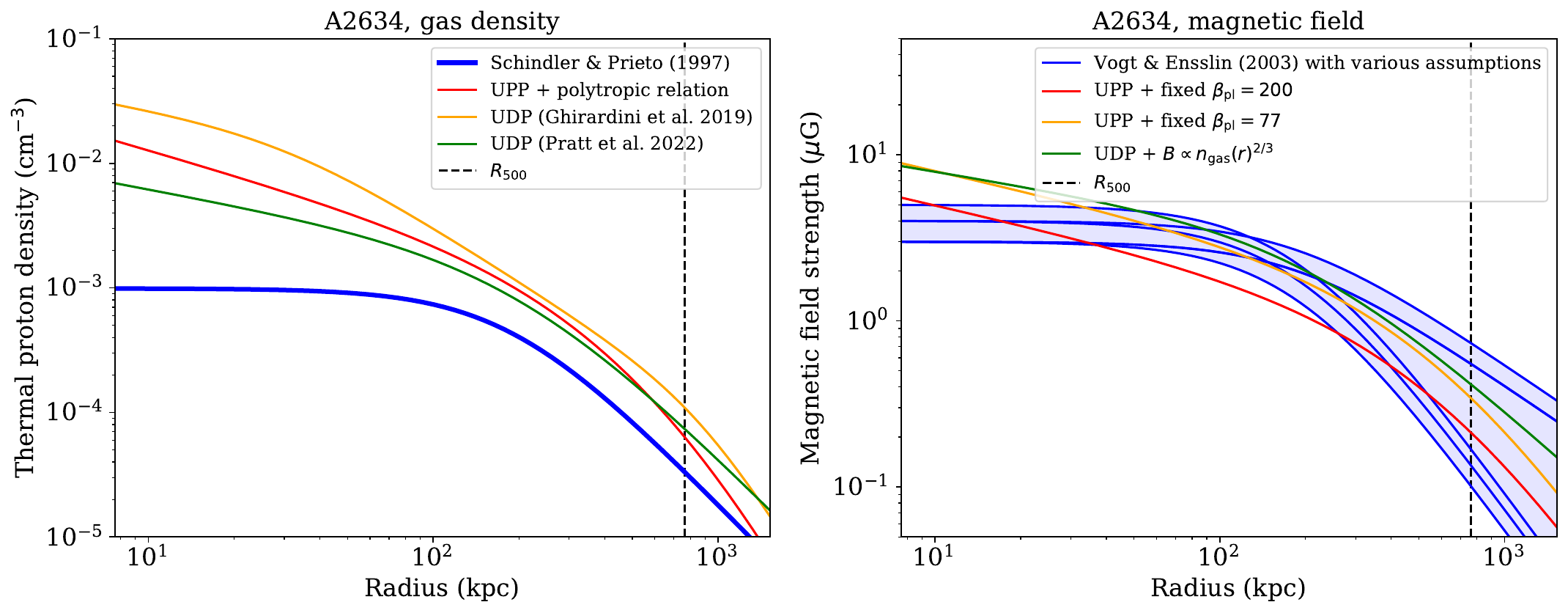}
\figsetgrpnote{A\,2364 cluster ($M \simeq 10^{14} \ M_{\odot}$): typical relaxed cluster with a dense core.}
\figsetgrpend

\figsetgrpstart
\figsetgrpnum{7.6}
\figsetgrptitle{Thermal-proton and magnetic-field densities of A119}
\figsetplot{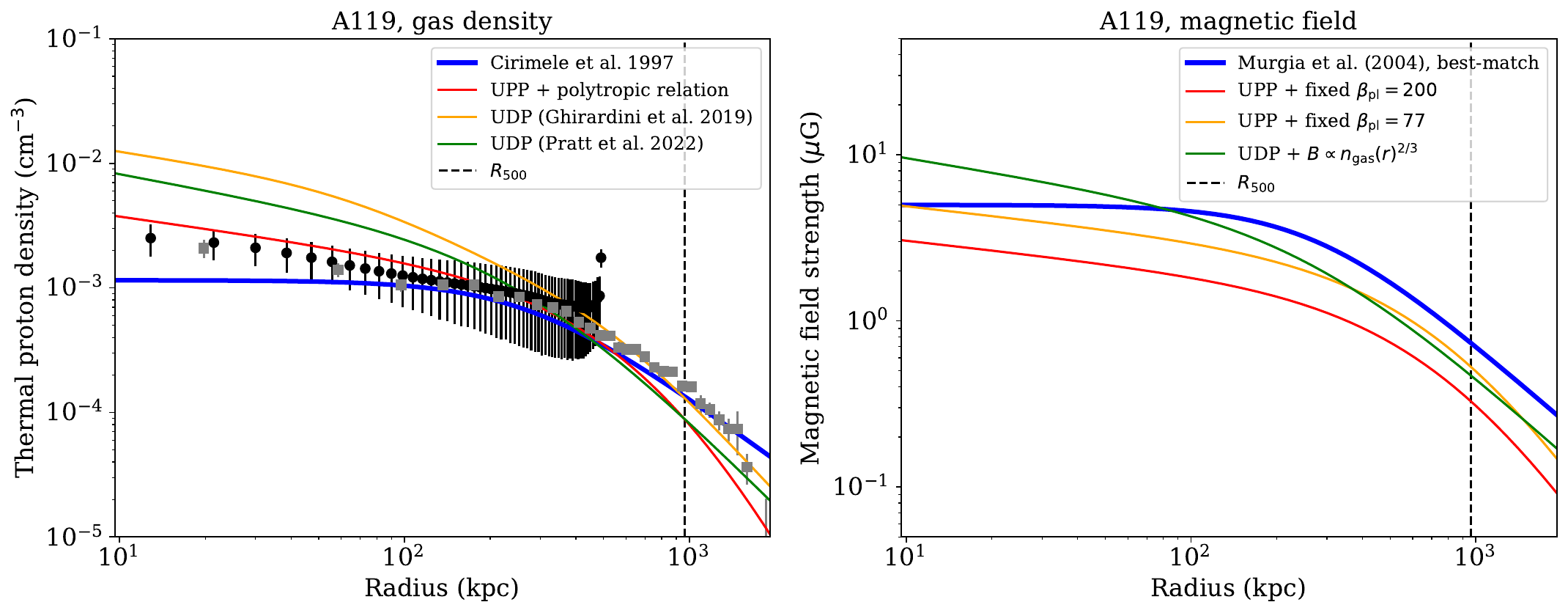}
\figsetgrpnote{A\,119 cluster ($M \simeq 2.5 \cdot 10^{14} \ M_{\odot}$): merging cluster with flatter core.}
\figsetgrpend

\figsetgrpstart
\figsetgrpnum{7.7}
\figsetgrptitle{Thermal-proton and magnetic-field densities of A400}
\figsetplot{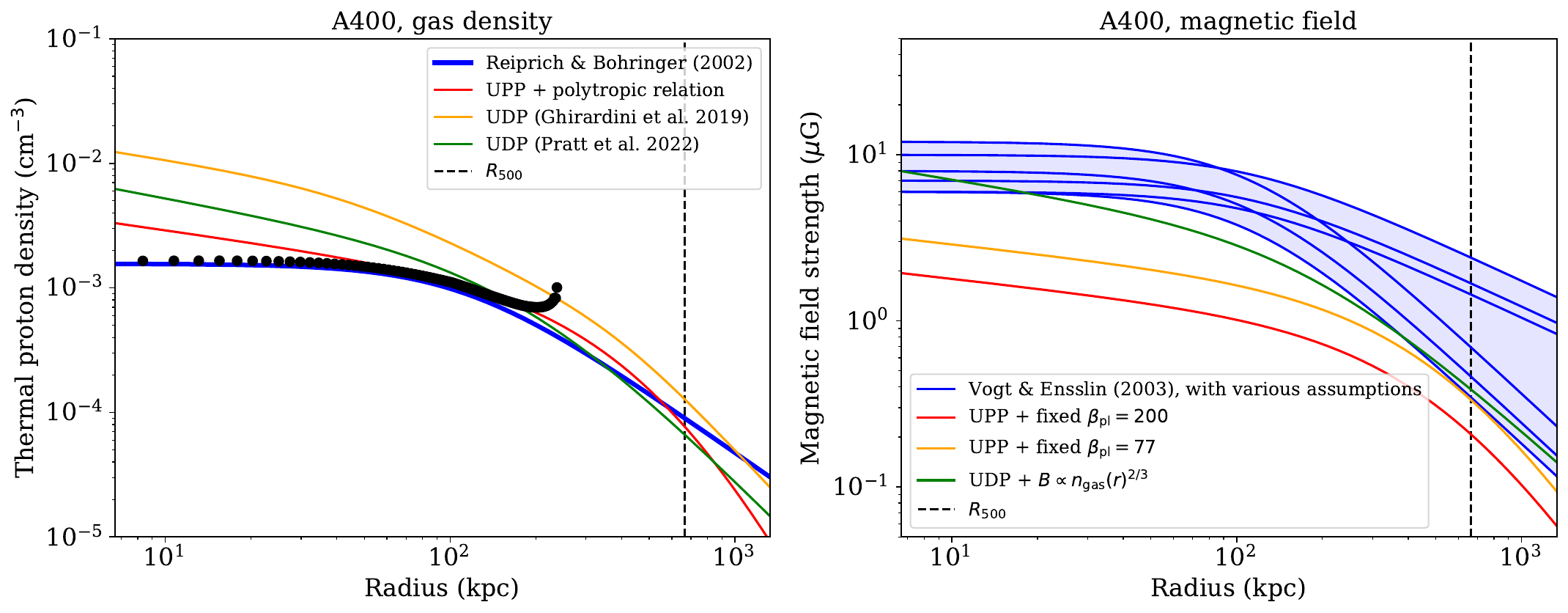}
\figsetgrpnote{A\,A400 cluster ($M \simeq 8 \cdot 10^{13} \ M_{\odot}$): merging cluster with flatter core.}
\figsetgrpend

\figsetgrpstart
\figsetgrpnum{7.8}
\figsetgrptitle{Thermal-proton and magnetic-field densities of A2199}
\figsetplot{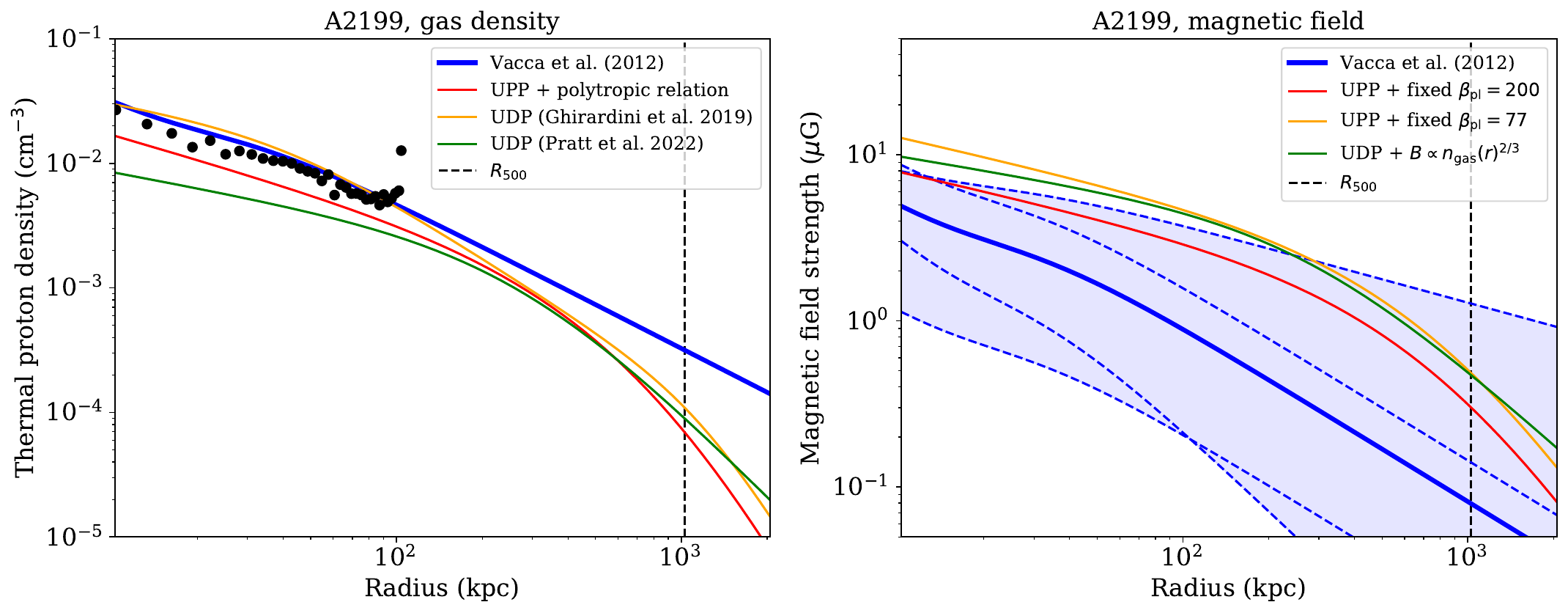}
\figsetgrpnote{A\,2199 cluster ($M \simeq 3 \cdot 10^{14} \ M_{\odot}$): typical relaxed cluster with a dense core..}
\figsetgrpend

\figsetgrpstart
\figsetgrpnum{7.9}
\figsetgrptitle{Thermal-proton and magnetic-field densities of A665}
\figsetplot{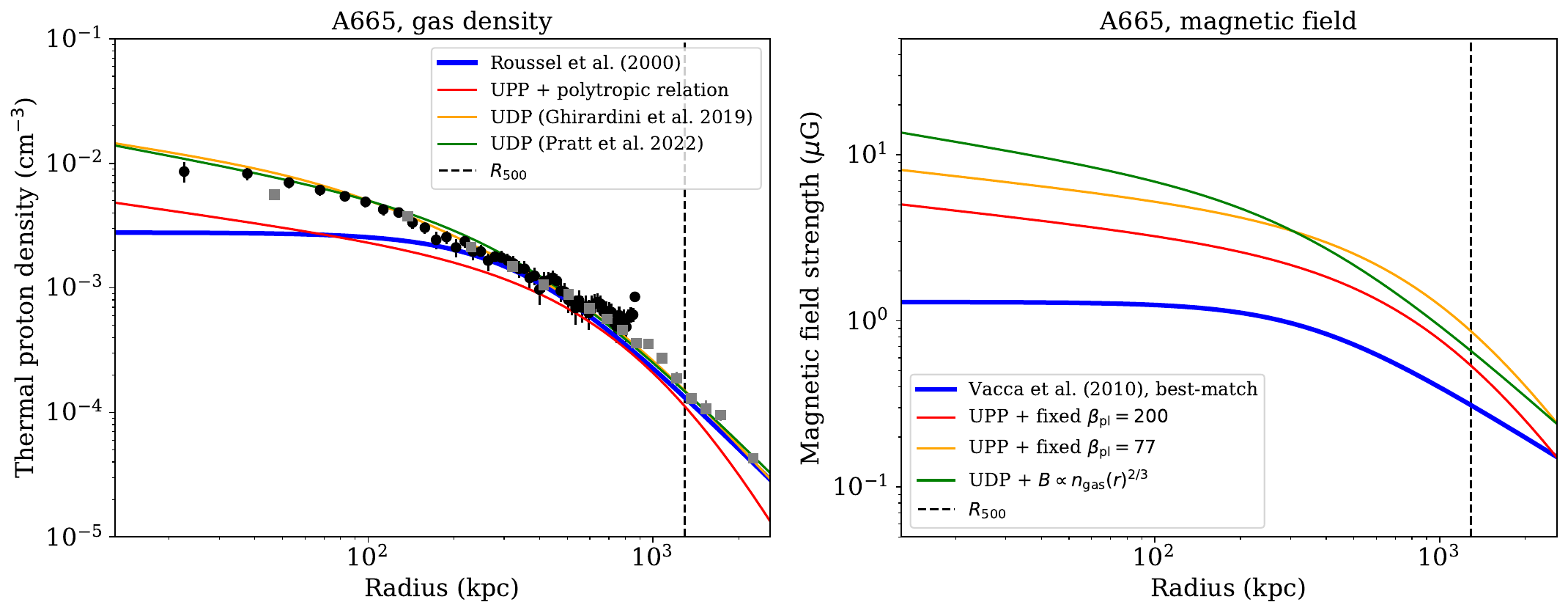}
\figsetgrpnote{A\,A665 cluster ($M \simeq 7 \cdot 10^{14} \ M_{\odot}$): merging cluster with flatter core.}
\figsetgrpend

\figsetgrpstart
\figsetgrpnum{7.10}
\figsetgrptitle{Thermal-proton and magnetic-field densities of A2255}
\figsetplot{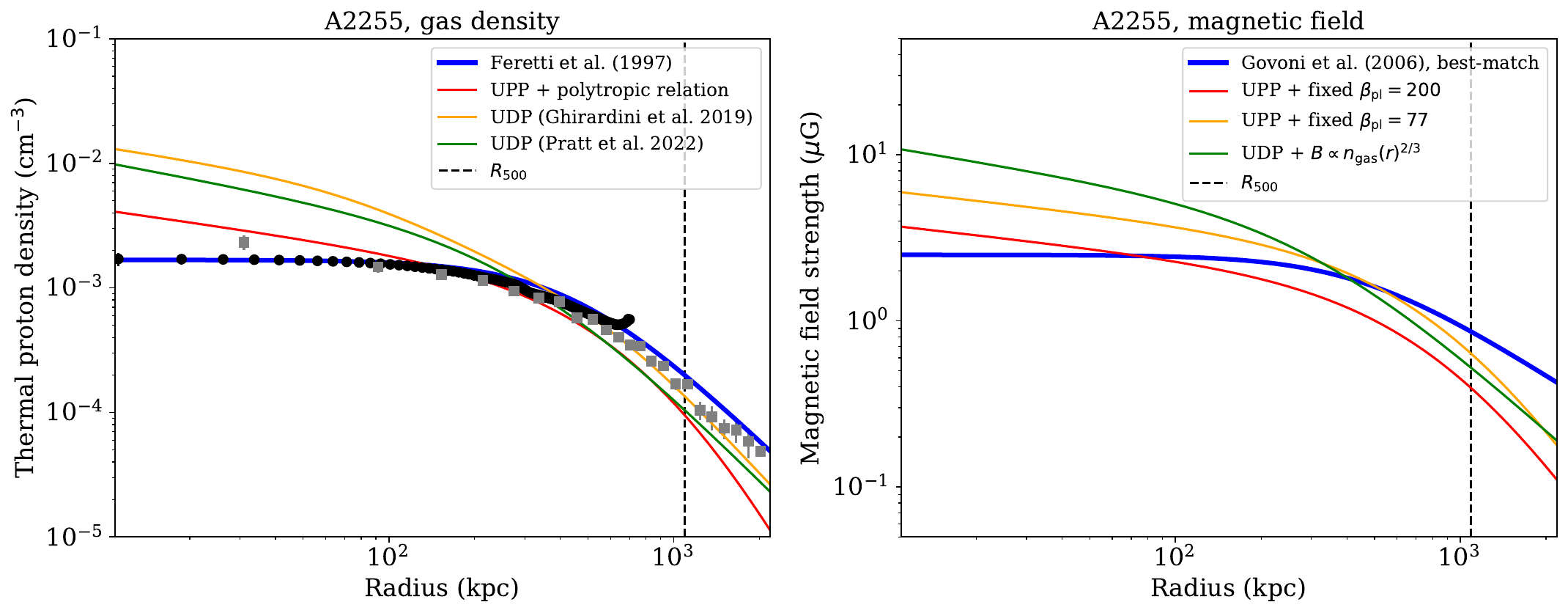}
\figsetgrpnote{A\,2255 cluster ($M \simeq 4 \cdot 10^{14} \ M_{\odot}$): merging cluster with flatter core.}
\figsetgrpend

\figsetgrpstart
\figsetgrpnum{7.11}
\figsetgrptitle{Thermal-proton and magnetic-field densities of A2382}
\figsetplot{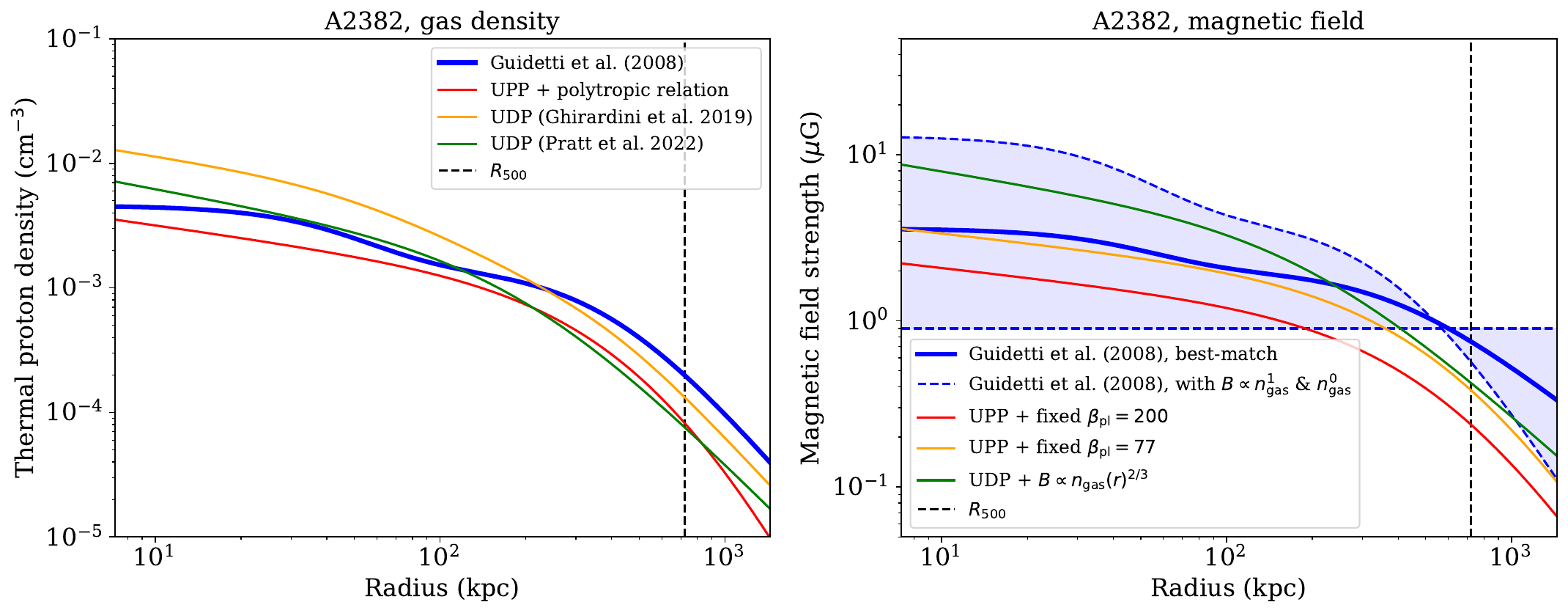}
\figsetgrpnote{A\,2382 cluster ($M \simeq 10^{14} \ M_{\odot}$): typical relaxed cluster with a dense core.}

\figsetend

%

\vspace{5mm}



\bibliography{sample631}{}
\bibliographystyle{aasjournal}



\end{document}